# Sign Reversal of anisotropic magnetoresistance in La$_{0.7}$Ca$_{0.3}$MnO$_3$/STO ultrathin films

Himanshu Sharma[1,a)], A. Tulapurkar[2], C. V. Tomy[1,b)]

[1]Department of Physics, Indian Institute of Technology Bombay, Powai, Mumbai – 400 076, India.
[2]Department of Electrical Engineering, Indian Institute of Technology Bombay, Powai, Mumbai – 400 076, India.

We present the observation of strain induced sign reversal of anisotropic magnetoresistance (AMR) in La$_{0.7}$Ca$_{0.3}$MnO$_3$ (LCMO) ultrathin films (thickness ~4 nm) deposited on SrTiO$_3$ (001) substrate (STO). We have also observed unusually large AMR (~24%) in LCMO/STO thin films with thickness of 6 nm below but close to its Curie temperature ($T_C$) which decrease as the film thickness increases. The sign reversal of AMR (with a maximum value of – 6 %) with magnetic field or temperature for the 4 nm thin film may be attributed to the increase in tensile strain in the plane of the thin film which in turn facilitates the rotation of the magnetization easy axis.

Anisotropic magnetoresistance (AMR) in manganites has attracted much attention due to its two-fold importance: (i) AMR properties can be used for exploring the physics due to the complex coupling between charge, spin and orbital states and (ii) possibility of its use devices like magnetic field sensors, low-cost inertial navigation systems and magnetoresistive random access memory (MRAM) [1-3], etc. Tremendous efforts have been devoted to obtain large values of AMR near room temperature. Spin-orbit coupling due to the double exchange interaction[4-7] is considered as one of the possible explanation for the origin of anisotropy in manganite thin films. It is possible to tune the AMR[18] either through intrinsic strains (e.g., doping, structural manipulations, strain induction, etc.) or through extrinsic strains (e.g., application of external electric or magnetic fields, temperature variation, etc.). Recently, several groups have reported strain induced AMR and the effect of strain on the metal to insulator transition $T_P$ in manganites thin films [4-10]. It is found that the othorhombic La$_{0.7}$Ca$_{0.3}$MnO$_3$ (LCMO) with an in-plane lattice constant 3.863 Å can be grown on SrTiO$_3$ (001) substrate (STO), which has the cubic structure with a lattice constant 3.905 Å such that the lattice mismatch between the STO and the LCMO is only ~1%. AMR has been found to be remarkably large in highly strained ultrathin films as compared to the AMR in thicker films[1,5,6,9-12] at certain temperatures below their ferromagnetic Curie temperature ($T_C$) which is attributed to the increase in the strain as the thickness of the film decreases[12, 13].

In this paper, anisotropy in epitaxial LCMO ultrathin films (4 nm, 6 nm and 8 nm) grown on STO (001) substrate is investigated. For a comparison, a relatively thicker film (~50 nm) is also investigated. Ultrathin films of LCMO down to 3 nm were deposited on STO substrates from a sintered target, using a KrF-Pulsed Laser Deposition (PLD) system with λ = 248 nm and energy density[7] of ~2 Jcm$^{-2}$. The typical lateral sizes of the LCMO films were 100 μm × 2 mm. The thin films were patterned with gold pads for electrical resistivity measurements (four probe) using optical lithography and ion beam milling process.

Magnetization of the thin films, as a function of temperature and applied magnetic field was measured using a SQUID-VSM (QD, USA). The temperature dependence of the field-cooled (FC) magnetization of the films with various thicknesses with an applied magnetic field of 0.1 T is shown in Fig. 1(a). For the comparatively strain relaxed thicker film of 50 nm, the $T_C$ occurs at a temperature close to the value reported for bulk LCMO (~248 K)[17]. However, it is observed that the $T_C$ decreases with decreasing thickness in LCMO films down to the critical thickness[13] of 3 nm ($T_C$ = ~106 K). Figure 1(b) shows derivatives of magnetization to highlight the shift in the ferromagnetic Curie temperature $T_C$. Inset of Fig. 1(b) shows the variations of $T_C$ (taken as the peak temperature in the derivative plot) as a function of film thickness.

It was also evident that the magnetization and hence the magnetic moment decreases with decreasing thicknesses in the films below the magnetic ordering temperature. This is further confirmed through the magnetization measurements as a function of magnetic field at 5 K, which is shown in Fig. 2 for films of thickness 3 nm, 4 nm, 6 nm and 8 nm. It is evident that even for the thinnest film (3 nm), we see a saturation magnetization, confirming the ferromagnetic nature of this film.

Figures 3(a), 3(b) and 3(c) show the variation of resistivity as a function of temperature for 50 nm, 6 nm and 4 nm thin films. In zero applied magnetic field, the metal to insulator transition (MIT) for the 50 nm thin film is observed near its $T_C$. As we increase the applied magnetic field, MIT temperature $T_P$ also increases, same as the $T_C$ behavior found in the magnetization. The shift in $T_P$ (in zero and applied fields) is also observed with the reduction in the thickness of the LCMO films down to 6 nm. The peak corresponding to $T_P$ disappears altogether for films of thickness 4 nm, as shown in Fig. 3(c). The MIT occurs only in applied fields for this film.

[a)]Email: himsharma@iitb.ac.in
[b)]Email: tomy@iitb.ac.in

Figure 3(d) shows the MR plots of thin films of 50 nm, 8 nm, 6nm and 4 nm, respectively. The percentage MR is calculated using the formula, $([\rho(0) - \rho(H)]/\rho(0)) \times 100$; where $\rho(0)$ and $\rho(H)$ (where H = 5 T ) are the resistivity in zero magnetic field and applied magnetic field of H, respectively. The largest MR is observed in the highly strained ultrathin films (4 nm) in comparison with the MR values in the relaxed thicker film of ~50 nm. More than 90% MR is observed in all the ultrathin films close to their transition temperature.

AMR of the thin films was measured by varying the angle from $0^o$ to $360^o$ at different temperatures for an applied magnetic field of 5 T, where $\theta = 0^o$ corresponds to the configuration of H perpendicular to the plane of the film and $90^o$ corresponds to H parallel to the plane of the film. The angular dependence of the resistivity for 50 nm, 8 nm and 6 nm thin films plotted in polar coordinates is shown in Fig. 4(a), 4(b) and 4(c), respectively. Corresponding linear plots of 50 nm, 8 nm, 6 nm along with that of 4 nm thin films are shown in figure 4(d), 4(e), 4(f) and 4(g), respectively (we could not use the log scale for the polar plot in the case of 4 nm thin film). The AMR percentage is defined as, $[\{\rho(\theta°) - \rho(90°)\}/\rho(90°)] \times 100$; where $\rho(90°)$ and $\rho(\theta°)$ are the resistivity corresponding to the configuration of H parallel to the plane of the film and H oriented at an angle $\theta$ with the plane, respectively.

Large values of AMR have been observed in highly strained ultrathin films of LCMO at 200 K (17% for 8 nm) and 100 K (24% for 6 nm), the temperatures being below but close to the $T_C$. On the other hand, a maximum AMR of only 7% is observed in the thicker film of 50 nm, which is the typical value obtained for less strained thin films[5]. The AMR for the LCMO thin films of 50 nm, 8 nm and 6 nm shows two-fold $\cos^2\theta$ dependence symmetry at all temperatures. However, in the 4 nm thin film, we observe a totally different behavior. Even though the AMR shows a two-fold $\cos^2\theta$ dependence at high temperatures, the AMR flips and changes sign as the temperature is lowered below $T_C$ (see Fig. 4(g)). The maximum of AMR (–6%) is observed at 100 K which decreases as the temperature decreases further.

Our observation of the sign change in the 4 nm LCMO thin film grown on STO substrate is entirely different to that of the observation by M. Egilmez et. al.,[9,11]. They could not observe a sign change in the AMR for the LCMO/STO films (down to 7 nm) even though they could observe a change in the sign of the AMR at temperatures below $T_C$ in the LCMO/LAO thin films. This difference in observation may be related to the fact that they have investigated thin films down to only 7 nm, whereas we have seen the effect in thin films of thickness 4 nm. We have also not observed any sign change in our film with a thickness of 6 nm, which is in agreement with the observation by M. Egilmez et. al. Figures 5(a) and 5(b) show the AMR obtained for the 6 nm and 4 nm thin films of LCMO/STO as function of applied magnetic field at 100 K. We do not see any sign change for the 6 nm thin film, whereas the sign change occurs for the 4 nm thin film as a function of applied field. The AMR has a $\cos^2\theta$ dependence at low fields, transforms into $\sin 2\theta$ dependence in the intermediate fields and then changes into $\sin^2\theta$ (changes sign) dependence when the magnetic field exceeds 3.5 T (see Fig. 5(b)). At the same time, the 6 nm thin film shows only a $\cos^2\theta$ dependence at all fields (see Fig. 5(a)). It is clear that the magnetic field also plays an important role in the switching of AMR sign, in addition to the thickness of the film. This is further confirmed in Figs. 5(c) and 5(d), where we show the maximum AMR (defined as $[\{\rho(0°) - \rho(90°)\}/\rho(90°)] \times 100$) for the 6 nm and 4 nm thin films at 100 K as a function of applied field. The field at which the switching of AMR sign (Fig. 5d) matches well with the filed in Fig. 5(b).

The results obtained in this paper in conjunction with the results from M. Egilmez et. al.,[9,11] bring in interesting aspects regarding the sign reversal of AMR. The same material, LCMO ultrathin films here, with different strains shows sign reversal in AMR at different thicknesses and different applied fields. With the compressive negative strain, the thickness needed is higher and the magnetic field needed is smaller as compared to that of the films with the tensile positive strain. Even though there is no clear understanding as to the reason for the reversal of AMR, one of the possibilities, especially in the case of manganite thin films, is the strain-induced rotation of the magnetization easy axis. This implies that larger magnetic fields and thinner films are needed to rotate the easy axis of the thin films with the tensile strain. High field or low field switching devices can be fabricated by designing these materials with tensile or compressive strains. It will be interesting to see whether such reversal in magnetization can also be observed directly by measuring the magnetization in the presence of applied gate voltage using insulating or ferroelectric gate.

In conclusion, we have shown that the sign reversal of AMR in LCMO ultrathin films is possible as a function of temperature and magnetic field once the film thickness is less than 6 nm. We have also observed enhanced AMR for ultrathin films which may be attributed to the enhancement of magnetic disorder due to strain in ultrathin films.

**ACKNOWLEDGMENTS**

We are grateful for availability of the Institute central facility (SQUID-VSM) in the Department of Physics and IITB Nanofabrication facility (IITBNF) in the Department of Electrical Engineering, Indian Institute of Technology Bombay.

[a)]Email: himsharma@iitb.ac.in
[b)]Email: tomy@iitb.ac.in

[a)] Email: himsharma@iitb.ac.in

[b)] Email: tomy@iitb.ac.in

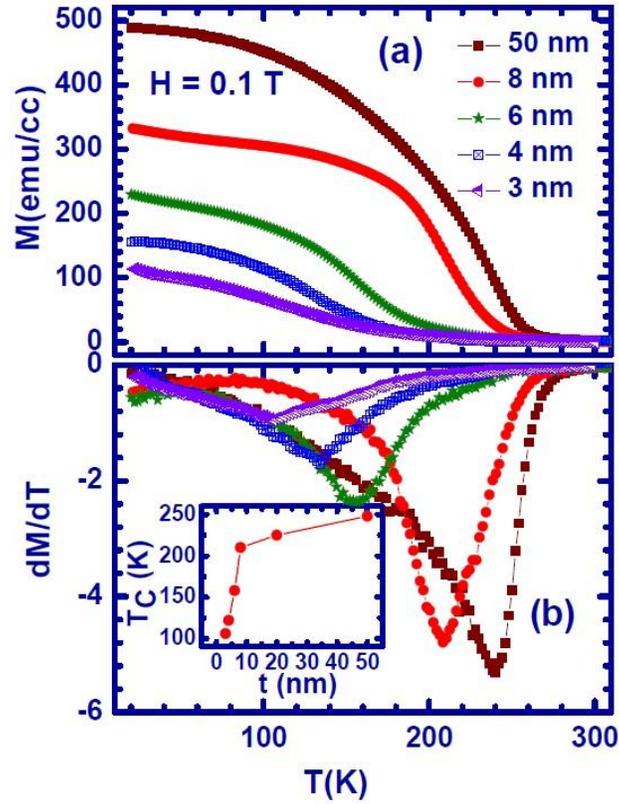

FIG. 1 (a) Field-Cooled (FC) magnetization as a function of temperature (T) in an applied field of 0.1 T for 3 nm, 4 nm, 6 nm, 8 nm and 50 nm LCMO/STO thin films. (b) The derivatives of magnetization to highlight the shift in $T_C$. Inset shows the variation of $T_C$ as a function of film thickness (t).

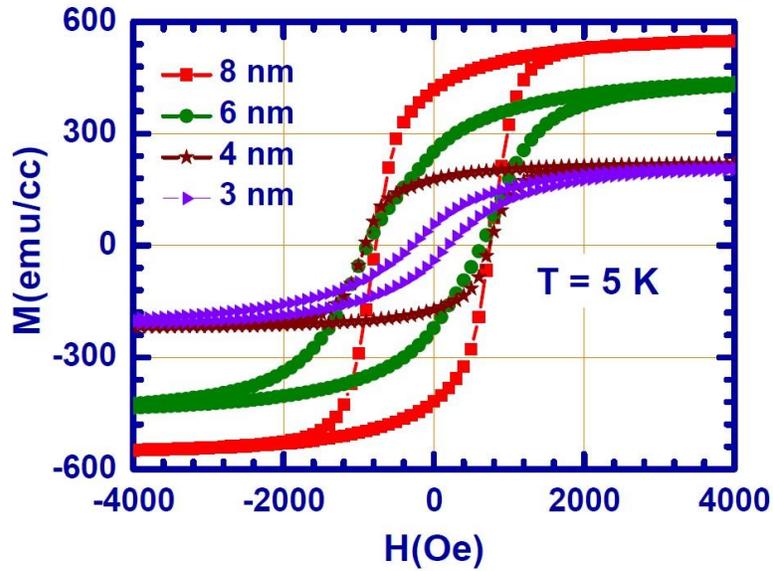

FIG. 2. Magnetization (M) curves of LCMO/STO ultra thin films at 5 K measured with magnetic field (H) parallel to the film plane.

[a)] Email: himsharma@iitb.ac.in
[b)] Email: tomy@iitb.ac.in

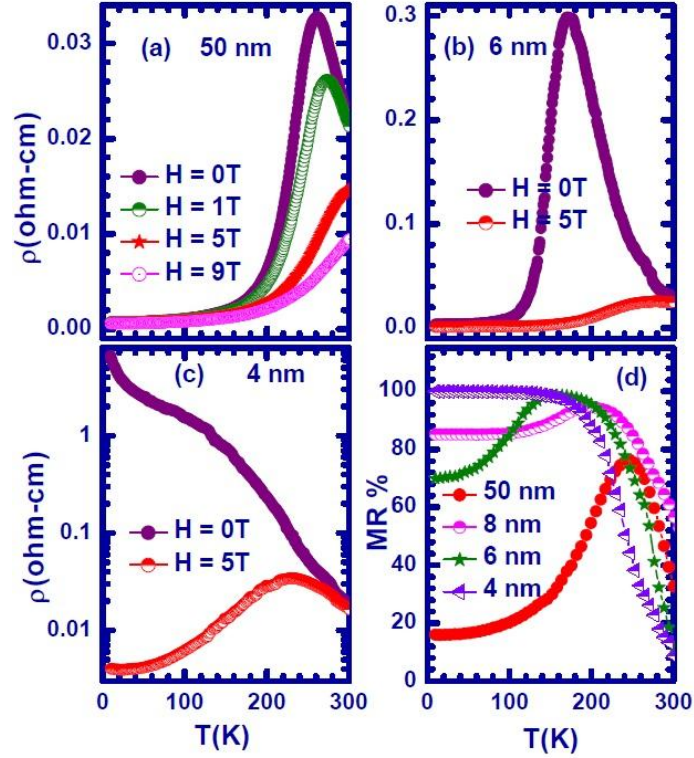

FIG. 3. Temperature dependence of the resistivity for LCMO/STO thin films of (a) 50 nm (b) 6 nm and (c) 4 nm (in log scale for 4 nm) in applied fields as given. (d) MR vs T plots of 50 nm, 8 nm, 6 nm and 4 nm thin films.

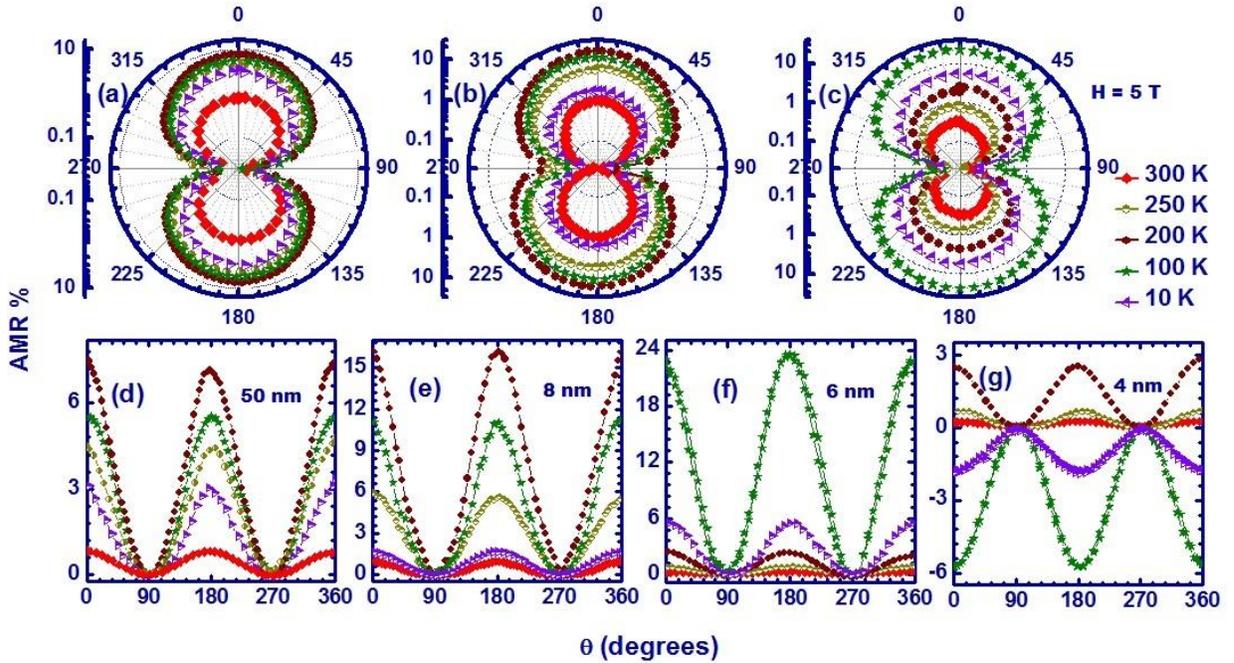

FIG. 4. Anisotropic Magnetoresistance (AMR) for LCMO/STO thin films of thickness (a) 50 nm, (b) 8 nm and (c) 6 nm, in applied field of 5 T, plotted in polar coordinates. AMR for thin films of thickness (d) 50 nm, (e) 8 nm, (f) 6 nm and (g) 4 nm plotted in linear coordinates with angle θ swept forward and backward (0°-360°-0°).

[a)]Email: himsharma@iitb.ac.in  
[b)]Email: tomy@iitb.ac.in

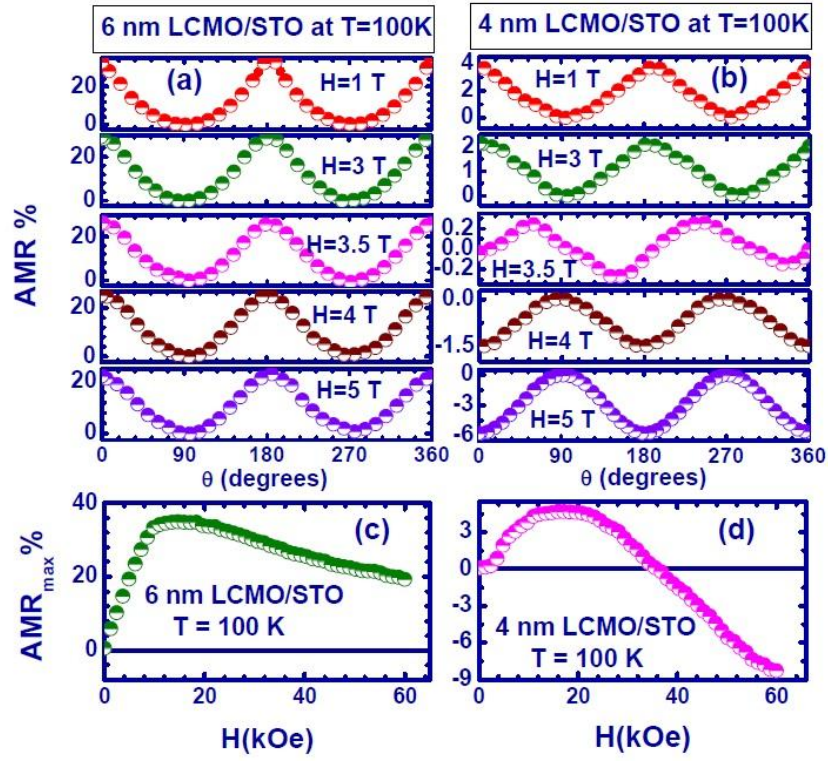

FIG. 5. AMR measured at 100 K for LCMO/STO films with thickness (a) 6 nm and (b) 4 nm as a function of angle (θ) in different magnetic fields. Maximum AMR vs applied field for (c) 6 nm and (d) 4 nm thin films.

[a] Email: himsharma@iitb.ac.in
[b] Email: tomy@iitb.ac.in